\documentclass[proceedings]{JHEP3}

\PrHEP{PrHEP hep2001}
\conference{International Europhysics Conference on HEP}

\usepackage{epsfig}
\usepackage{amsmath}
\usepackage{amssymb}

\newcommand{\beq}{\begin{equation}}
\newcommand{\eeq}{\end{equation}}

\def\dg{\Delta \Gamma_d}
\def\bea{\begin{eqnarray}}
\def\eea{\end{eqnarray}}

\def\barr{\begin{eqnarray}}
\def\earr{\end{eqnarray}}

\def\lsim{\raise0.3ex\hbox{$\;<$\kern-0.75em\raise-1.1ex\hbox{$\sim\;$}}}
\def\gsim{\raise0.3ex\hbox{$\;>$\kern-0.75em\raise-1.1ex\hbox{$\sim\;$}}}

\def\gh{\Gamma_H}
\def\gl{\Gamma_L}
\def\dg{\Delta \Gamma_d}
\def\aa{{\cal A}}

\title{The Width Difference of $B_d$ Mesons}

\author{Amol Dighe\\
   Max-Planck-Institute for Physics, F\"ohringer Ring 6,
        D-80805 Munich, Germany}              
\author{\speaker{Tobias Hurth}\\
CERN, Theory Division, CH--1211 Geneva 23, Switzerland} 
\author{Choong Sun Kim\\
      Department of Physics and IPAP, Yonsei University, 
Seoul 120-749, Korea     } 
\author{Tadashi Yoshikawa \\
Department of Physics,
          University of North Carolina,
        Chapel Hill, NC 27599-3255, USA}

\abstract{
We estimate $\dg/\Gamma_d$, including $1/m_b$ contributions
and part of the next-to-leading order QCD corrections.
We find that adding the latter corrections decreases 
the value of $\dg/\Gamma_d$ computed at the leading order by a factor of 
almost $2$. 
We also show that under certain conditions an 
upper bound on the value of $\dg/\Gamma_d$
in the presence of new physics can be derived.
With the high statistics and accurate time resolution of the upcoming
LHC experiment, the measurement of $\dg$ seems to be possible.
This measurement would be important for an accurate 
measurement of $\sin(2\beta)$ at the LHC. 
In addition, we point out the possibility that 
the measurement of  the width difference leads to a clear signal for new physics.\\
\vspace{0.5cm} CERN-TH/2001-333, \,\, MPI-PhT/2001-49, \,\, IFP-802-UNC 
}

\begin{document}
%
%

  \section{Introduction}
    \label{sec:intro}

The two mass eigenstates of the neutral $B_d$ system 
have slightly different lifetimes.
Within the standard model (SM), the difference in the decay 
widths, however, is CKM-suppressed with respect to 
that in the $B_s$ system. A rough estimate leads to
$\frac{\Delta \Gamma_d}{\Gamma_d} \, \sim \, 
\frac{\Delta \Gamma_s}{\Gamma_s} \cdot \lambda^2
\approx  0.5 \% ~~,$\, 
where $\lambda = 0.225$ is the sine of the Cabibbo angle, and we have taken 
$\Delta \Gamma_s/\Gamma_s \approx 15\%$ \cite{BBGLN} (see also \cite{Ben,Bec}). 
Here $\Gamma_{d(s)} = (\Gamma_L + \Gamma_H)/2$ is the average decay width 
of the  light and heavy $B_{d(s)}$ mesons ($B_L$ and $B_H$ respectively).
We denote these decay widths by $\Gamma_L,
\Gamma_H$ respectively,
and define $\Delta \Gamma_{d(s)} 
\equiv \gl - \gh$. 

At the present accuracy of measurements,
this lifetime difference $\dg$ can well be ignored. As a result,
the measurement and the phenomenology of $\dg$ have been neglected
so far, as compared with the lifetime difference in the $B_s$ system
for example. However,
with the possibility of experiments with high time resolution and 
high statistics, such as at the  LHC, 
this quantity is becoming more and more relevant.

Taking the effect of $\dg$ into account is important in two
aspects. There is the interlinked nature of the accurate
measurements of $\beta$ and $\dg/\Gamma_d$ through the
conventional gold-plated decay. 
In the future experiments that aim to measure $\beta$ to an 
accuracy of 0.005 or better, the corrections due to
$\dg$ will form an important  part of the systematic error.
On the other hand, the measurement of $\dg$ allows for the possibility
to detect a clear signal for new physics beyond the SM.

It is known that, if $(\Gamma_{21})_s$ is unaffected by new
physics, the value of $\Delta \Gamma_s$ in the $B_s$ 
system is bounded from above by its value as calculated in the
SM. In the $B_d$ system, this statement does not strictly hold true. 
However, if  $(\Gamma_{21})_d$ is unaffected by new physics and
the unitarity of the $3\times 3$ CKM matrix holds, 
an upper bound on the value of $\dg$ may then be found.

With the possibility of experiments with high time
resolution and high statistics, it is worthwhile to have a look
at this quantity and make a realistic estimate of the 
possibility of its measurement (see also \cite{Paper}).

\section{Measurability of $\dg$}
    \label{sec:measure}

At LHCb, the proper time resolution is expected to be
as good as $\Delta \tau \approx 0.03$ ps.
This indeed is a very small fraction of the $B_d$ lifetime
($\tau_{B_d} \approx 1.5$ ps \cite{pdg}), 
so the time resolution
is not a limiting factor in the accuracy of the measurement,
 and  the statistical error plays the dominant role.
Taking into account the estimated
number of $B_d$ produced --- for example the number of 
reconstructed $B_d \rightarrow J/\psi \, K_S$ events at the LHC is expected
to be $5 \times 10^5$ \, (\cite{lhc} table 3) ---  the measurement of the
lifetime difference does not look too hard at first glance.
One may infer that if the number of relevant events
with the proper time of decay measured with the
precision $\Delta \tau$ is $N$, then the value
of $\Delta \Gamma_d / \Gamma_d$ is measured with an
accuracy of $1 / \sqrt{N}$. With a
sufficiently large number of events $N$, it should be
possible to reach the accuracy of 0.5\%
or better.

However, the time measurements of the decay 
of an untagged $B_d$ to a single final state can only be sensitive to
quadratic terms in $\dg/\Gamma_d$. This would imply that, for determining 
$\dg/\Gamma_d$ using only one final state, the accuracy of the
measurement needs to be $(\dg/\Gamma_d)^2 \sim  10^{-5}$.
 In \cite{Paper}  we gave an explicit derivation of that 
general  statement, pointing out the exact conditions under which 
the above statement is valid. Ways of getting around these conditions
lead us to the decay modes that can provide measurements sensitive
linearly to $\dg/\Gamma_d$.  
This discussion indicates the necessity of combining measurements from two
different final states in order to be sensitive to a quantity
 that is linear in $\dg/\Gamma_d$.

A viable option, perhaps the most efficient among the ones 
considered in \cite{Paper}, is to compare the measurements of the
untagged lifetimes  of the semileptonic decay mode $\tau_{SL}$ and 
of the CP-specific decay modes $\tau_{CP_{\pm}}$.
The ratio between the two lifetimes 
$\tau_{CP\pm}$ and $\tau_{SL}$ is 
\begin{equation}
\frac{\tau_{SL}}{\tau_{CP\pm}} = 1 \pm \frac{\cos(2\beta)}{2} 
\frac{\dg}{\Gamma_d}
+ {\cal O} \left[ (\dg/\Gamma_d)^2 \right]
~~.
\label{cp-sl}
\end{equation}
The measurement of these two lifetimes 
should be able to give us a value of $|\dg|$,
since $|\cos(2\beta)|$ will already be known to a good accuracy
by that time.

Since the CP-specific decay modes of $B_d$ 
(e.g. $J/\psi K_{S(L)}, D^+ D^-$) have smaller
branching ratios than the semileptonic modes, and 
the semileptonic data sample may be enhanced 
by including the self-tagging decay modes (e.g. $D_s^{(*)+}D^{(*)-}$)
 which also have large branching ratios,
we expect that the most useful combination will be the
measurement of $\tau_{SL}$ through all self-tagging decays and
that of $\tau_{CP_+}$ through the decay $B_d \to J/\psi K_S$.
After 5 years of LHC running, 
we should have about $5 \times 10^5$ events of 
$J/\psi K_S$, whereas the number of semileptonic decays,   
at LHCb alone, that will be directly useful in the lifetime 
measurements is expected to be more than $10^6$ per year, 
even with conservative estimates of efficiencies.

\section{Estimation of $\dg$}
    \label{sec:analysis}

In \cite{Paper} we estimated $\dg/\Gamma_d$ including $1/m_b$ contributions
and part of the next-to-leading order QCD corrections.
We find that adding the latter corrections decreases 
the value of $\dg/\Gamma_d$ computed at the leading order by 
a factor of almost $2$. The final result is
\begin{equation}
\Delta \Gamma_d / \Gamma_d = (2.6 ^{+1.2}_{-1.6}) \times 10^{-3} \, .
\end{equation}
Using another expansion of the partial NLO QCD corrections proposed
in \cite{breport}, we get
\begin{equation}
\Delta \Gamma_d / \Gamma_d = (3.0 ^{+0.9}_{-1.4}) \times 10^{-3}\, ,
\end{equation}
where we have used the preliminary values 
for the bag factors from the JLQCD collaboration~\cite{jlqcd}. 
In the error estimation, the errors are
the uncertainties on the values of
the CKM parameters, of the bag parameters, of the mass of the $b$ quark,
and of the measured value of $x_d$.
Further sources of error are the assumption of naive factorization 
made for  the $1/m_b$ matrix elements, 
the scale dependence  and the missing  terms in the NLO contribution. 
Although the latter error is decreased 
in the second estimate by smallness of  CKM factors, 
a complete NLO calculation is definitely desirable for the 
 result to be more reliable.

\section{Interlinked Nature of  $\sin(2 \beta)$ and $\dg$}
\label{gold}

The time-dependent CP asymmetry measured through the 
``gold-plated'' mode $B_d \to J/\psi K_S$ is
\begin{equation}
\aa_{CP}  = \frac{\Gamma[\bar{B}_d(t) \to J/\psi K_S] - 
\Gamma[B_d(t) \to J/\psi K_S]}
{\Gamma[\bar{B}_d(t) \to J/\psi K_S] +
\Gamma[B_d(t) \to J/\psi K_S]} \, \approx   \sin(\Delta m_d t) \sin(2\beta)~~,
\label{acp-approx}
\end{equation}
which is valid when the lifetime difference, the direct CP
violation, and the mixing in the neutral $K$ mesons are 
neglected. As the accuracy of this measurement increases,
the corrections due to these factors will have to be taken 
into account. Keeping only linear terms in small quantities,
we obtain 
\barr
\aa_{CP} & = & \sin(\Delta m t) 
        \sin(2\beta) \left[ 1 - \sinh \left( \frac{\dg t}{2} \right)
        \cos(2 \beta) \right] 
\label{dg-corr} \\
& & + 2 {\rm Re}(\bar{\epsilon}) 
\left[ -1 + \sin^2(2 \beta) \sin^2(\Delta m t) - \cos(\Delta m t) 
\right]
\label{re-eps-corr} \\
& & + 2 {\rm Im}(\bar{\epsilon}) \cos(2\beta) \sin(\Delta m t) ~~.
\label{im-eps-corr} 
\earr
The first term in (\ref{dg-corr}) represents the  standard approximation 
used (\ref{acp-approx}) and the correction due to the 
lifetime difference $\dg$.
The rest of the terms [(\ref{re-eps-corr}) and (\ref{im-eps-corr})]
are corrections due to the CP violation in $B$--$\bar{B}$ and
$K$--$\bar{K}$ mixings. Note that $\bar{\epsilon}$ is an effective parameter
that absorbs several small uncertainties and equals a few
$\times 10^{-3}$~(see \cite{Paper}).

The BaBar collaboration gives the bound on the coefficient
of $\cos(\Delta m t)$ in (\ref{re-eps-corr}), while neglecting the other
correction terms \cite{babar-direct}. 
When the measurements are accurate enough to
measure the $\cos(\Delta m t)$ term, 
the complete
expression for ${\cal A}_{CP}$ above (\ref{dg-corr}--\ref{im-eps-corr})
needs to be used.
In the future experiments 
that aim to measure $\beta$ to an accuracy of 0.005
\cite{lhc}. The corrections due to $\bar{\epsilon}$ and 
$\dg$ will form a major part of the systematic error, which
can be taken care of by a simultaneous fit to
$\sin(2\beta), \dg$ and $\bar{\epsilon}$.

\section{New Physics}
\label{contrast}

The calculations of the lifetime difference in $B_d$ 
and in the $B_s$ system (as in \cite{BBGLN}) run along
similar lines. However, there are some subtle differences involved,
due to the values of the different CKM elements involved, which
have significant consequences. 

In particular, whereas the upper
bound on the value of $\Delta\Gamma_s$ (including the effects of
new physics) is the value of $\Delta\Gamma_s({\rm SM})$ \cite{grossman},
the upper bound on $\Delta\Gamma_d$ involves a multiplicative
factor in addition to $\Delta\Gamma_d({\rm SM})$:
using the definitions 
$\Theta_q \equiv {\rm Arg}(\Gamma_{21})_q , 
\Phi_q \equiv {\rm Arg}(M_{21})_q $,
where $q \in \{ d,s \}$, we can write 
\begin{equation}
\Delta\Gamma_q = - 2 |\Gamma_{21}|_q  \cos(\Theta_q - \Phi_q)~~.
\label{theta-phi}
\end{equation}
Since the contribution to $\Gamma_{21}$ comes only from tree
diagrams, we expect the effect of new physics on this quantity 
to be very small. We therefore take
$|\Gamma_{21}|_q$ and $\Theta_q$ to be unaffected by new physics. 
On the other hand, the mixing phase $\Phi_q$ appears from loop
diagrams and can therefore be very sensitive to new physics.
Based on these assumptions, one derives an upper bound on new physics
within the $B_s$ system \cite{grossman}:
\begin{equation}
\Delta\Gamma_s \leq \frac{\Delta \Gamma_s({\rm SM})}{\cos(2\Delta\gamma)}
\approx \Delta \Gamma_s({\rm SM})~~,
\label{dgs-bd}
\end{equation}
with $ 2\Delta\gamma
\approx - 0.03 $.
Thus, the value of $\Delta \Gamma_s$ can only decrease in the
presence of new physics.

In the $B_d$ system, an upper bound for $\dg$,  
based on the additional assumption of three-generation
unitarity, can be derived: 
\begin{equation}
\Delta \Gamma_d \leq  \frac{\Delta \Gamma_d({\rm SM})}
{\cos[ {\rm Arg}(1 + \delta f) ]}~~.
\label{dgd-bd}
\end{equation}
We can calculate the
bound (\ref{dgd-bd}) in terms of the extent of the higher order NLO
corrections. In \cite{Paper}, we got $|{\rm Arg}(1 + \delta f)| < 0.6$,
so that we have the bound  $\Delta \Gamma_d < 1.2 ~ \Delta \Gamma_d({\rm SM})$. 
A complete NLO calculation will be able to give a
stronger bound.

We have seen that the ratio of 
two effective lifetimes can enable us to measure the quantity
$\Delta\Gamma_{obs(d)} \equiv \cos(2\beta) \Delta\Gamma_d/\Gamma_d$.
In the presence of new physics, this quantity is in fact
(see eq.~(\ref{theta-phi}))
$\Delta\Gamma_{obs(d)} =  - 2 (|\Gamma_{21}|_d/\Gamma_d)  
\cos(\Phi_d) \cos(\Theta_d - \Phi_d)$. In SM, we get 
\begin{equation}
\Delta\Gamma_{obs(d)}({\rm SM}) = 2 (|\Gamma_{21}|_d/\Gamma_d) 
\cos(2 \beta) \cos[ {\rm Arg}(1 + \delta f) ]~~.
\end{equation}
If $|\delta f| < 1.0$, we have $\cos[{\rm Arg}(1+\delta f)] > 0$ 
(in fact, from the fit in \cite{ckm-fit} and our error estimates, 
we have $\cos[{\rm Arg}(1+\delta f)] > 0.8$). 
Then $\Delta\Gamma_{obs(d)}({\rm SM})$ is predicted to be positive.
New physics is not expected to affect $\Theta_d$, but it may affect 
$\Phi_d$ in such a way as to make the combination 
$\cos(\Phi_d) \cos(\Theta_d - \Phi_d)$ change sign.
A negative sign of $\Delta\Gamma_{obs(d)}$ would  therefore 
be a clear signal of such new physics.

It is well known, that the 
$B_d$--$\bar{B}_d$ mixing phase $\Phi_d$ is efficiently
measured through the decay modes $J/\psi K_s$ and $J/\psi K_L$. 
If we take the new physics effects into account, the time-dependent
asymmetry is  
${\cal A}_{CP} = - \sin(\Delta M_d t) \sin(\Phi_d)$; 
in the SM, we have $\Phi_d = -2\beta$. 
The measurement of $\sin(\Phi_d)$ still allows for a discrete ambiguity
$\Phi_d \leftrightarrow \pi - \Phi_d$. 
It is clear that,  
if $\Theta_d$ can be determined  
independently of the mixing in the $B_d$ system,  
then measuring $\Delta\Gamma_{obs(d)}$,  which is proportional to 
$\cos(\Phi_d) \cos(\Theta_d - \Phi_d)$,  
resolves the discrete ambiguity in principle. 
We note that these features are unique to the $B_d$ system.

\end{document}